# Optimization of MIMO detectors: Unleashing the multiplexing gain

Nirmalendu Bikas Sinha[1], S.Chakraborty[1], P. K. Sutradhar[1], R.Bera[2], And M.Mitra[3]

**Abstract**— Multiple Input Multiple Output (MIMO) systems have recently emerged as a key technology in wireless communication systems for increasing both data rates and system performance. There are many schemes that can be applied to MIMO systems such as space time block codes, space time trellis codes, and the Vertical Bell Labs Space-Time Architecture (V-BLAST). This paper proposes a novel signal detector scheme called MIMO detectors to enhance the performance in MIMO channels. , we study the general MIMO system, the general V-BLAST architecture with Maximum Likelihood (ML), Zero- Forcing (ZF), Minimum Mean-Square Error (MMSE), and Ordered Successive Interference Cancellation (SIC ) detectors and simulate this structure in Rayleigh fading channel. Also compares the performances of MIMO system with different modulation techniques in Fading and AWGN channels. Base on frame error rates and bit error rates, we compare the performance and the computational complexity of these schemes with other existence model.Simulations shown that V-BLAST implements a detection technique, i.e. SIC receiver, based on ZF or MMSE combined with symbol cancellation and optimal ordering to improve the performance with lower complexity, although ML receiver appears to have the best SER performance-BLAST achieves symbol error rates close to the ML scheme while retaining the low-complexity nature of the V-BLAST.

——————————— ◆ ———————————

## 1. INTRODUCTION

Future wireless communication networks will need to support extremely high data rates in order to meet the rapidly growing demand for broadband applications such as high quality audio and video. Existing wireless communication technologies cannot efficiently support broadband data rates, due to their sensitivity to fading.

Recent research on wireless communication systems has shown that using MIMO at both transmitter and receiver offers the possibility of wireless communication at higher data rates, enormous increase in performance and spectral efficiency compared to single antenna systems. The information-theoretic capacity of MIMO channels was shown to grow linearly with the smaller of the numbers of transmit and receiver antennas in rich scattering environments, and at sufficiently high signal-to-noise (SNR) ratios [1].MIMO wireless systems are motivated by two ultimate goals of wireless communications: high-data-rate and high-performance [2],[3].

During recent years, various space-time (ST) coding schemes have been proposed to collect spatial diversity and/or achieve high rates. Among them, V-BLAST (Vertical Bell Labs Layered Space-Time) transmission has been widely adopted for its high spectral efficiency and low implementation complexity [4]. When maximum-likelihood (ML) detector is employed, V-BLAST systems also enjoy receives diversity, but the decoding complexity is exponentially increased by the number of transmit-antennas. Although some (near-) ML schemes (e.g., sphere-decoding (SD), semi-definite programming (SDP)) can be used to reduce the decoding complexity, at low signal to-noise ratio (SNR) or when a large number of transmit antennas and/or high signal constellations are employed, the complexity of near-ML schemes is still high. Some suboptimal detectors have been developed, e.g., successive interference cancellations (SIC), decision feedback equalizer (DFE), which are unable to collect receive diversity [5]. To further reduce the complexity, one may apply linear detectors such as zero-forcing (ZF) and minimum mean-square error (MMSE) equalizers. It is well-known that linear detectors have inferior performance relative to that of ML detector. However, unlike ML detector, the expected performance (e.g., diversity order) of linear equalizers has not been quantified directly. The mutual information of ZF equalizer has been studied in [6] with channel state information at the transmitter.

————————————————

- *Prof. Nirmalendu Bikas Sinha, corresponding author is with the Department of ECE and EIE , College of Engineering & Management, Kolaghat, K.T.P.P Township, Purba- Medinipur, 721171, W.B., India.*
- *S.Chakraborty is with the Department of ECE, College of Engineering & Management, Kolaghat, K.T.P.P Township, Purba- Medinipur, 721171, W.B., India.*
- *P. K. Sutradhar is with the Department of ECE, College of Engineering & Management, Kolaghat, K.T.P.P Township, Purba- Medinipur, 721171, W.B., India.*
- *Dr. R. Bera is with the S.M.I.T, SikkimManipal University, Majitar, Rangpo, East Sikkim, 73713.*
- *Dr. M.Mitra is With the Bengal Engineering and science University, Shibpur, Howrah, India .*

In this paper, we propose a modified V-BLAST system, which introduces different delay offsets for each substreme in the transmitter. At the receiver, we can employ ZF strategy to recover information and the introduction of delay offsets enables the requirement of $Nr$ to be relaxed to $Nr \geq 1$ (in the conventional V-BLAST, $Nr \geq Nt$. Where, $Nr$ and $Nt$ are the receiver and transmitter antennas respectively. We will verify the performance improvement by theoretical analysis and simulation results. From our analysis, with ZF decoding, the diversity order can reach $Mr$ in the modified V-BLAST system. But the increase of the diversity order is at the cost of the multiplexing gain.

The main goal of our paper is to study the MIMO detectors schemes and quantify the diversity orders collected by linear equalizers for V-BLAST. Also optimize the ultimate detector and modulation technique that yields a better error performance than general V-BLAST.

The rest of this paper is organized as follows. In Section 2, the MIMO system model is introduced. Section 3 gives the performances of MIMO system with different modulation techniques in Fading and AWGN channels and Section 4 gives the performance analysis of the linear equalizers optimize the ultimate detector.

## 2. MIMO SYSTEM MODEL

Consider a multi-antenna system with $M$ transmit-antennas and $N$ receive-antennas. Furthermore, as a commonly used structure for the MIMO system, V-BLAST shares some basic modules with our general multiple an

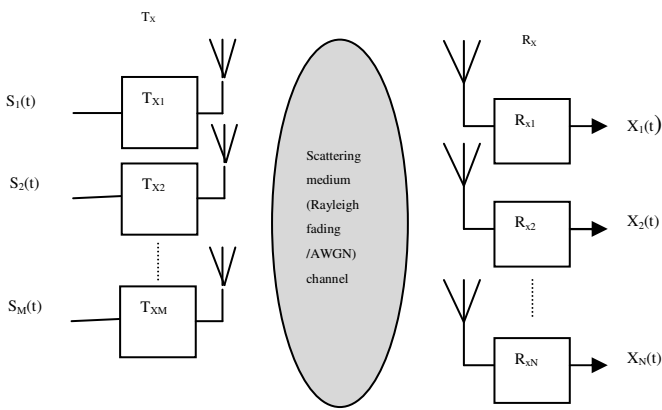

Fig.1: Diagrams of a single user MIMO communication system using M transmit and N receives antennas.

Let M and N be the number of transmit and receive antennas, respectively. The received signal in the $i^{th}$ antenna is given by $Y_i = \sum_{j=1}^{N} h_{ij} x_j + n_i$ ……. (1)

where i = 1, 2, 3…N, $h_{ij}$ is the fading corresponding to the path from transmit antenna j to receive antenna i. $n_i$ is the noise corresponding to receive antenna i.

$$y = \begin{bmatrix} y_1 \\ y_2 \\ \vdots \\ y_N \end{bmatrix} \ldots (2), \quad x = \begin{bmatrix} x_1 \\ x_2 \\ \vdots \\ x_M \end{bmatrix} \ldots (3)$$

$$n = \begin{bmatrix} n_1 \\ n_2 \\ \vdots \\ n_N \end{bmatrix} \ldots (4), \quad H = \begin{bmatrix} h_{11} & \cdots & h_{1M} \\ h_{21} & & h_{2M} \\ \vdots & \ddots & \vdots \\ h_{N1} & \cdots & h_{NM} \end{bmatrix} \ldots (5)$$

The above wireless channel is modulated as y=Hx+n. Where H is the channel matrix and n is the channel noise. For transmit/receive beam forming with the diversity of order MN, is considered as full diversity. On the other hand the antenna gain is; Max = {M,N} ≤ antenna gain ≤ MN

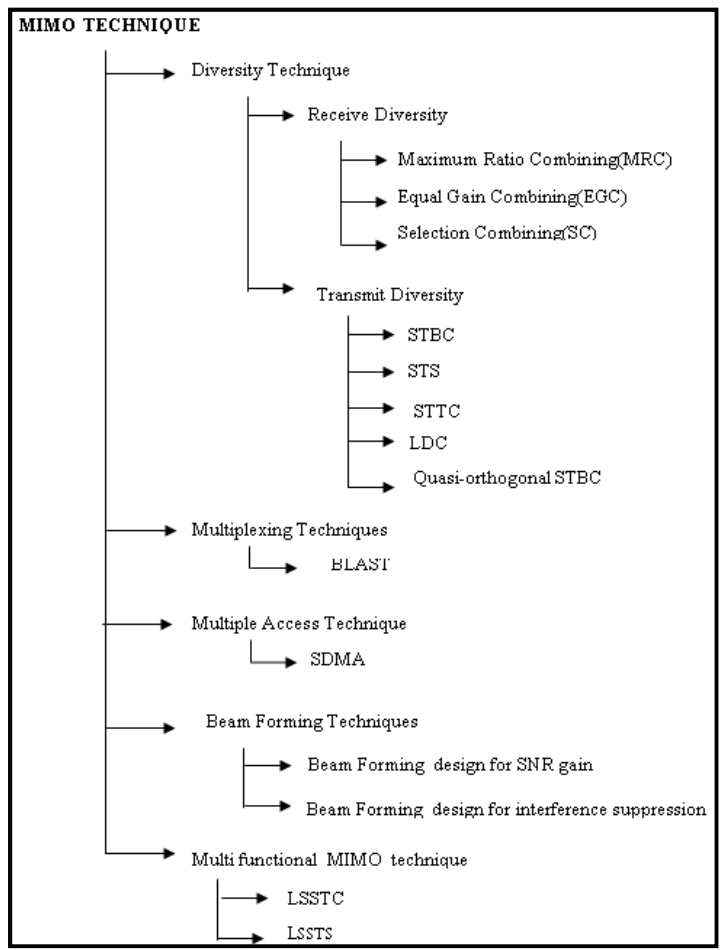

## 3. PERFORMANCES OF MIMO SYSTEM WITH DIFFERENT PROPAGATION CONDITION

Communication in a slow flat Rayleigh fading channel with AWGN is not reliable as the channel frequently enters into deep fades, (i.e., the channel attenuation is large). More specifically, as seen in class, Rayleigh fading converts an exponential dependency of bit error probability on the signal-to-noise ratio (SNR) into an inverse relationship. For BPSK, QPSK and 16 QAM the symbol error rate in an AWGN channel is

$$P_{BPSK} = Q\left(\sqrt{\frac{2E_m}{\sigma_n^2}}\right)\dots(6), \quad P_{QPSK} = Q\left(\sqrt{\frac{E_m}{\sigma_n^2}}\right)\dots\dots(7)$$

$$P_{16QAM} = \frac{3}{4}Q\left(\sqrt{\frac{E_m}{10\sigma_n^2}}\right)\dots\dots(8)$$

Where, $Q(x) = \frac{1}{2} erfc(x/1.414) \dots\dots\dots(9)$

However, with Rayleigh fading the average probability of bit error is

$$P_{e(BPSK)} = \frac{1}{2}\cdot\left(1-\sqrt{\frac{\beta}{(\beta+1)}}\right)\dots(10), \quad P_{e(QPSK)} = \frac{1}{2}\cdot\left(1-\sqrt{\frac{\beta}{(\beta+2)}}\right)\dots(11)$$

$$P_{e(16QAM)} = \frac{3}{4}\cdot\left(1-\sqrt{\frac{\beta}{(\beta+40)}}\right)\dots(12), \quad \beta = \frac{E_b}{N_0}\cdot E(\alpha^2)\dots..(13)$$

Using the model (Fig.1), the M ×N MIMO communication has M*N channels. This leads to the expressions for the received signal-to-noise ratios (SNRs) of each of these communications in terms of the MIMO channel.

$$(SNR)_{MIMO} = \frac{1}{2}(H)^2\frac{E_m}{\sigma_n^2}\dots\dots(14)$$

Where, $H = \sum_{i=1}^{M}\sum_{j=1}^{N}|h_{ij}|\dots\dots(15)$

Let $p(\theta,\varphi)$ be the probability density function of MIMO system using BPSK are given by,

$$P_{MIMO} = \int_0^{2\pi}\int_0^{\pi}Q\left(\sqrt{(H)^2\frac{E_b}{N_0}}\right)P(\theta,\varphi)\sin\theta d\theta d\varphi\dots(16)$$

Where, $h_{ij}$ are the channels that are set-up in the MIMO communication system and $\alpha$ is Rayleigh distribution function, $E_m$ is the signal power while $\sigma_n^2$ is the noise power spectral density.

Rician and Rayleigh [If $A/\varphi^2 = 0(Noise\ alone)$] probability density functions respectively.

$$f(r) = \frac{r}{\varphi^2}exp\left(-\frac{r^2}{2\varphi^2}\right)\dots\dots\dots(17)$$

When ($A/\varphi^2$) is very large Eq.(8) becomes a Gaussian probability density function of mean A and variance $\varphi^2$.

$$f(r) \approx \frac{1}{\sqrt{2\pi\varphi^2}}exp\left[-\frac{(r-A)^2}{2\varphi^2}\right]\dots\dots\dots(18)$$

Fig.2 shows the plots for the Rayleigh and Gaussian densities. Where μ is the mean, and σ is the standard deviation.

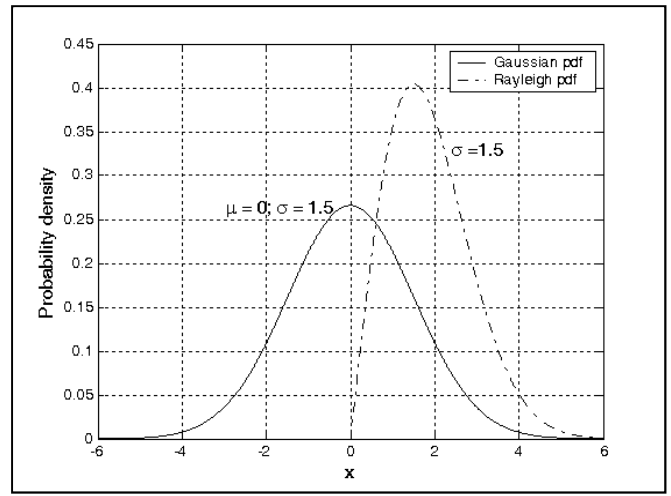

Fig.2: Gaussian and Rayleigh probability densities.

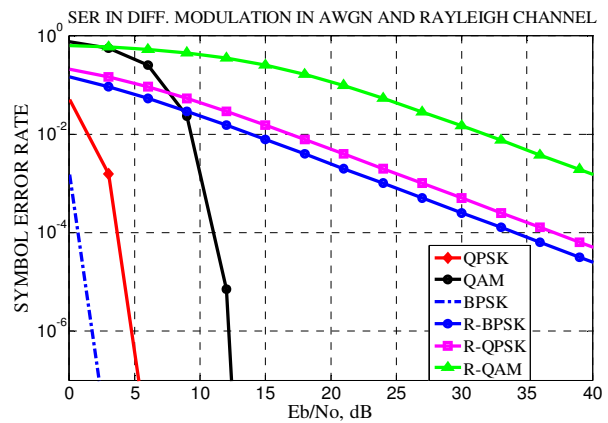

Fig.3 : SER for different modulation in AWGN and Rayleigh channel.

Fig.3 shows clearly this severe degradation in the probability of bit error for BPSK in a Rayleigh fading channel for $E(\alpha^2) = 1$. As it can be seen, the performance degradation is as severe as in the BPSK case. The overall system performance of MIMO system is better than other systems for different modulations due to its improvement of transmit and receive diversity. It is noticeable that the overall system performance of QPSK is better compare to other modulation schemes. For all SNR levels, MIMO system has the least SER, and hence enhance the diversity gain because the lower the error in the received signals, the higher is the detection. But the overall system performance of Rayleigh channel is better in comparison with other channels with respect to probability density function as well as SER vs. SNR.

## 4. AVERAGE BER ANALYSIS

4.1 *Maximum Likelihood (ML):*

Theoretically maximum likelihood (ML) detection algorithm is the optimum method of recovering the transmitted signal at the receiver. ML receiver is a method that compares the received signals with all possible transmitted signal vector which is modified by channel matrix H and estimates transmit symbol vector x according to the Maximum Likelihood principle , which is shown as:

$$\hat{x} = \arg_{x_k \in \{x_1, x_2 \ldots x_N\}} \min \|r - Hx_k\|^2 \ldots \ldots (19)$$

where, $x$ is the estimated symbol vector. Although ML detection offers optimal error performance, it suffers from complexity issues. It has exponential complexity in the sense that the receiver has to consider $|A|^M$ possible symbols for an M transmitter antenna system with A is the modulation constellation.

4.2 *V-BLAST Zero Forcing (ZF) ZF characteristic:*

Simple linear receiver with low computational complexity and suffers from noise enhancement. It works best with high SNR. The solution of the ZF is given by:

$$\hat{x} = (H^*H)^{-1}Hx = H^+x \ldots (20)$$ , Where $(\ )^+$ represents the pseudo-inverse.

4.3 *V-BLAST Minimum Mean Square Error (MMSE):*

The MMSE receiver suppresses both the interference and noise components, whereas the ZF receiver removes only the interference components. This implies that the mean square error between the transmitted symbols and the estimate of the receiver is minimized. Hence, MMSE is superior to ZF in the presence of noise. Some of the important characteristics of MMSE detector are simple linear receiver, superior performance to ZF and at Low SNR, MMSE becomes matched filter. Also at high SNR, MMSE becomes Zero-Forcing. MMSE receiver gives a solution of:

$$\hat{x} = D \cdot x = \left(\frac{1}{SNR}I_{N_R} + H^HH\right)^{-1} \cdot H^Hx \ldots (21)$$

At low SNR, MMSE becomes ZF:

$$\left(\frac{1}{SNR}I_{M_R} + H^HH\right)^{-1} H^H \approx \frac{1}{SNR}H^H \ldots (22)$$

At high SNR, MMSE becomes ZF:

$$\hat{x} = D \cdot x = \left(\frac{1}{SNR}I_{M_T} + H^HH\right)^{-1} H^H \approx (H^HH)^{-1}H^H \ldots (23)$$

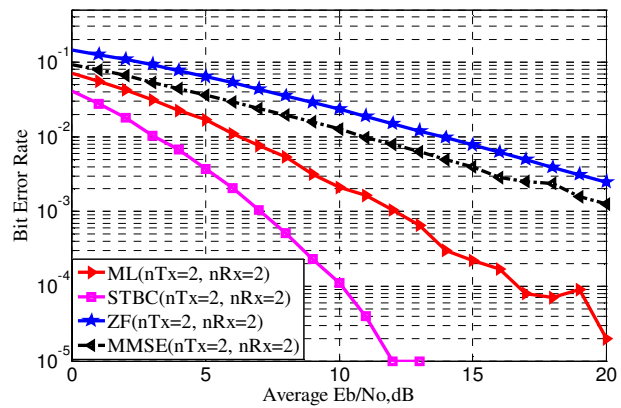

Figure 4, Performance curves for different linear detectors (ML,STBC, ZF, MMSE ) in 2×2 MIMO - V-BLAST system in a slow fading channel.

## CONCLUSION

In this paper, we analyze the performance of linear detectors for MIMO V-BLAST systems in slow fading channels for different modulations and different channels, which exhibit the best trade-off between performance and complexity among spatial multiplexing techniques. We show that conventional linear equalizers can only collect diversity N r – N t +1 for MIMO V-BLAST systems though they have very low complexity.By investigating and simulating each receiver concepts, it was shown that V-BLAST implements a detection technique, i.e. SIC receiver, based on ZF or MMSE combined with symbol cancellation and optimal ordering to improve the performance, although ML receiver appears to have the best SER performance. In this paper, the MIMO principle is based on a rich multipath environment without a normal Line-of-Sight (LOS) that is the Rayleigh flat fading channel, due to movement or other changes in the environment, LOS situation can arise. So finally we proposed that ML detector for MIMO-V-Blast in slow fading channel with QPSK modulation is the ultimate optimization technique in the next generation broadband communication system.

## ACKNOWLEDGEMENT

Authors would like to thank to the contribution of Makar Chand Snai , Sumit Kumar Gupta and Manish Sonal pursuing B.Tech in the Department of Electronics & Communication Engineering at College of Engineering and Management, Kolaghat, under WBUT in 2010, W.B, India for their extensive hard work and sincere support in completing this project.

**Prof. Nirmalendu Bikas Sinha** received the B.Sc (Honours in Physics), B. Tech, M. Tech degrees in Radio-Physics and Electronics from Calcutta University, Calcutta,India,in1996,1999 and 2001, respectively. He is currently working towards the Ph.D degree in Electronics and Telecommunication Engineering at BESU. Since 2003, he has been associated with the College of Engineering and Management, Kolaghat. W.B, India where he is currently an Asst.Professor is with the department of Electronics & Communication Engineering & Electronics & Instrumentation Engineering. His current research Interests are in the area of signal processing for high-speed digital communications, signal detection, MIMO, multiuser communications,Microwave /Millimeter wave based Broadband Wireless Mobile Communication ,semiconductor Devices, Remote Sensing, Digital Radar, RCS Imaging, and Wireless 4G communication. He has published large number of papers in different international Conference, proceedings and journals.He is currently serving as a member in international journal editorial board and reviewer for Wireless communication and radar system in different international journals.

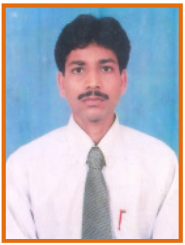

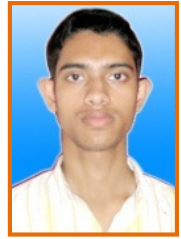

**Sourav Chakraborty** is pursuing B.Tech in the Department of Electronics & Communication Engineering at College of Engineering and Management, Kolaghat, under WBUT in 2011, West Bengal, India. His areas of interest are in Microwave /Millimeter wave based Broadband Wireless Mobile Communication and digital electronics. He has published some papers in different international journals.

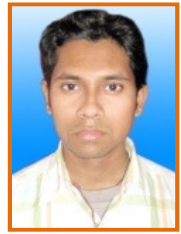

**Prosenjit Kumar Sutradhar** is pursuing B.Tech in the Department of Electronics & Communication Engineering at College of Engineering and Management, Kolaghat, under WBUT in 2011, West Bengal, India. His areas of interest are in Microwave /Millimeter wave based Broadband Wireless Mobile Communication and digital electronics.He has published some papers in different international journals.

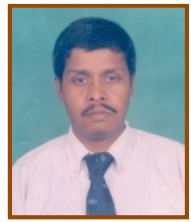

**Dr. Rabindranath Bera** is a professor and Dean (R&D), HOD in Sikkim Manipal University and Ex-reader of Calcutta University, India. B.Tech, M.Tech and Ph.D.degrees from Institute of Radio-Physics and Electronics, Calcutta University. His research areas are in the field of Digital Radar, RCS Imaging, Wireless 4G Communications, Radiometric remote sensing. He has published large number of papers in different national and international Conference and journals.

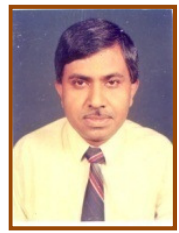

**Dr. Monojit Mitra** is an Assistant Professor in the Department of Electronics & Telecommunication Engineering of Bengal Engineering & Science University, Shibpur. He obtained his B.Tech, M.Tech & Ph. D .degrees from Calcutta University. His research areas are in the field of Microwave & Microelectronics, especially in the fabrication of high frequency solid state devices like IMPATT. He has published large number of papers in different national and international journals. He has handled sponsored research projects of DOE and DRDO. He is a member of IETE (I) and Institution of Engineers (I)society.